\documentclass[twocolumn,aps,showpacs,prb,tightenlines,amsmath,amssymb]{revtex4}
\usepackage{bm}
\usepackage{graphicx}              
\usepackage{amssymb}               
\usepackage{amsmath}                     
\usepackage{colordvi}
\usepackage{calrsfs} \newcommand{\bgreek}[1]{\mbox{\boldmath$#1$\unboldmath}}
\begin{document}   

\title{Theory of optical excitation spectra and depolarization dynamics in
  bilayer WS$_2$ from viewpoint of excimers}
\author{T. Yu}
\author{M. W. Wu}
\thanks{Author to whom correspondence should be addressed}
\email{mwwu@ustc.edu.cn.}
\affiliation{Hefei National Laboratory for Physical Sciences at
  Microscale and Department of Physics, 
University of Science and Technology of China, Hefei,
  Anhui, 230026, China} 
\date{\today}

\begin{abstract}
We investigate the optical excitation spectra and the photoluminescence
depolarization dynamics in bilayer WS$_2$. A different understanding of
 the optical excitation spectra in the recent photoluminescence experiment
by Zhu {\em et al.} [arXiv:1403.6224] in bilayer WS$_2$ is proposed. In the
experiment, four excitations (1.68, 1.93, 1.99 and
2.37~eV)
  are observed and identified to be indirect exciton for the $\Gamma$ valley,
trion, A exciton
and B exciton excitations, respectively, with the redshift for the 
A exciton energy
measured to be 30$\sim$50~meV when the sample synthesized from
 monolayer to bilayer.
According to our study, by considering there exist both the intra-layer and
charge-transfer excitons in the bilayer WS$_2$, with inter-layer 
hopping of the hole, there exists
excimer state composed by the superposition of the intra-layer and
charge-transfer exciton states. Accordingly, we show that the four optical
excitations in the bilayer WS$_2$
are the A charge-transfer exciton,
${\rm A}'$ excimer, ${\rm B}'$ excimer and B
 intra-layer exciton states, respectively, with the calculated 
resonance energies showing good agreement with the
experiment. In our picture, the speculated indirect exciton, which involves 
a high-order phonon absorption/emission process, is not necessary. 
Furthermore, the binding energy for the excimer state is calculated
to be 40~meV, providing reasonable explanation for the experimentally observed energy redshift of the A
exciton. Based on the excimer states, we further derive the exchange interaction Hamiltonian. 
 Then the photoluminescence depolarization dynamics due to the
 electron-hole exchange interaction is studied in the pump-probe
setup by the kinetic spin Bloch
equations. We find that there is
always a residual photoluminescence polarization that is exactly half of the
  initial one, lasting
for an extremely long time, which is robust against the
initial energy broadening and strength of the momentum
scattering. This large steady-state photoluminescence polarization
indicates that the photoluminescence relaxation time is extremely
long in the steady-state photoluminescence experiment, and can be the
cause of the anomalously large photoluminescence polarization, nearly
100$\%$ observed in the experiment by Zhu {\em et al.} in the bilayer
WS$_2$. This steady state is shown to come from the
unique form of the exchange interaction Hamiltonian, under which
 the density matrix evolves into
the one which commutes with the
exchange interaction Hamiltonian.

\end{abstract}

\maketitle 

\section{Introduction} 
\begin{figure}[htb]
  {\includegraphics[width=8.7cm]{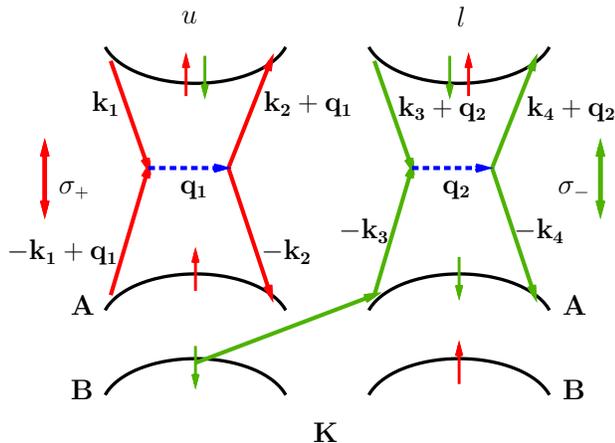}}
  \caption{(Color online) Schematic diagram of the exciton configurations and
    the intra-valley exchange interaction processes for the bright exciton in
    the K valley for BL TMDs. In the figure, $u$ and $l$ represent ``upper'' and ``lower''
    layer, respectively, which is associated with the absorption of $\sigma_{+}$ and $\sigma_{-}$
    light directly in the K valley. A and B denote the A and B IL
    (CT) exciton if the e-h pair, which is labeled by the same
    color, is in the same $u$/$l$ (different) layer.
    The Feynman diagrams
    show the two dominant exciton exchange
    interaction progresses, with momentum conservation explicitly shown in the figure.
 On one hand, the e-h pair in one IL exciton can virtually recombines
and then generates another IL exciton due to the Coulomb interaction
directly, shown by the red arrows. On the other hand, there exists another higher-order process, in which
the hole in the CT exciton first hops from one layer to another
and then recombines virtually with the electron part to generate the
IL exciton due to the Coulomb interaction, shown by the green arrows.}
\label{figyw1}
\end{figure}

In the past several years, as a new type of two-dimensional material,
monolayer (ML) transition metal dichalcogenides (TMDs) have attracted much
attention partly 
 due to their novel optical properties arising from their
 unique band structures.\cite{MoS_1,MoS_2,
valley_wang18,valley_wang20,valley_wang24,directgap_wang11,
direct_gap3,directgap_wang23,direct_gap5,direct_gap6,
splitting_wang27,splitting_wang31,splitting_wang37,splitting_wang38,absorption_Mak,absorption_Kioseoglou,absorption_wang15} With the
direct energy gap and large energy splitting of
 the valence bands,\cite{MoS_1,MoS_2,directgap_wang11,
direct_gap3,directgap_wang23,direct_gap5,direct_gap6,
valley_wang18,valley_wang20,valley_wang24} the chiral optical
valley selection rule allows the optical control of the valley and spin
degrees in ML TMDs,
 which are mainly realized by the excitonic excitation.\cite{MoS_2,valley_wang20,absorption_Mak,valley_wang24,
absorption_Kioseoglou,valley_wang18,absorption_wang15,absorption_Marie}
 Consisting of two ML TMDs, with the added layer degree of freedom, bilayer (BL) TMDs
 also exhibit rich optical properties due to the preservation of the chiral optical
valley selection rule,\cite{absorption_Mak,indirect,electric1,Hamiltonian1,Hamiltonian2,electric2,long_live,inter-layer,Cui,WS1} apart from the new features such as the electrical polarization, electrical-tuned
magnetic moments and magnetoelectric effect.\cite{electric1,Hamiltonian1,Hamiltonian2,electric2} Specifically,
due to the added layer degree of freedom in BL TMDs, the optical-excited electron and hole can not only stay in the same
 layer, which form the intra-layer (IL) exciton, but also in different layers
 referred to as the charge-transfer (CT) exciton [the configurations for the A
   and B IL (CT) excitons in the K valley are shown in
   Fig.~\ref{figyw1}]. Furthermore, due to the inter-layer coupling, the
 two kinds of excitons can couple to form a new elementary excitation:
 excimer.\cite{excimer1,excimer2} Therefore, BL TMDs may provide an ideal
 platform to study the
 excimer optical excitation and related photoluminescence (PL)
 depolarization dynamics.

 Very recently, a great deal of attention has been drawn to BL TMDs from theoretical and
 experimental
 aspects.\cite{Hamiltonian1,Hamiltonian2,absorption_Mak,indirect,electric1,electric2,long_live,inter-layer,Cui}
 The theoretical studies show that only the hole with the same spin in the
   same valley can hop between different
 layers efficiently in BL
  TMDs.\cite{Hamiltonian1,Hamiltonian2} However, it is further claimed that 
due to the inter-layer hopping energy of the hole is smaller than the energy
splitting of the valence bands, the inter-layer hopping of the hole is markedly
suppressed and hence 
  there may exist spin-layer locking effect in BL TMDs.\cite{Hamiltonian2} In
  this sense, BL TMDs
can be treated as two separated ML TMDs, which has been used to understand the
recent experiments related to the optical exciton excitation and PL depolarization dynamics.\cite{indirect,electric1,Hamiltonian2,electric2,long_live,inter-layer,Cui}

Experimentally, the optical exciton spectra and related PL depolarization
  dynamics in BL TMDs are in active 
 progress.\cite{absorption_Mak,indirect,electric1,Hamiltonian1,Hamiltonian2,electric2,long_live,inter-layer,Cui} 
The recent PL
experiments in BL TMDs show that the spectra of the optical excitation is
very  
different from the ML situation.\cite{indirect,long_live,inter-layer,Cui,WS1} On
one hand, in the BL
 TMD heterostructures, excitation energy much lower than the one in
 ML TMDs is observed and attributed to be CT exciton, whose
 lifetime is found to be as long as nanoseconds.\cite{long_live,inter-layer}
 On the other hand, in the experiments for the BL
 WS$_2$ 
 carried out by Zhu {\em et al.},\cite{Cui, WS1} it
has been observed that there are four resonance excitations with excitation
energies approximately being 1.68, 1.93, 1.99 and
2.37~eV, respectively, rather than the two excitations named A and
B excitons with resonance energies approximately being 2.03 and 2.40~eV in
the ML WS$_2$. These
four excitations are speculated to be the indirect exciton for the $\Gamma$ valley,
trion, A exciton
and B exciton excitations, respectively.\cite{Cui, WS1} Specifically, compared
to the ML WS$_2$, the obvious redshift for the A exciton energy about
 30 $\sim$ 50~meV is observed in the BL WS$_2$ in these experiments.\cite{Cui, WS1} Moreover, in the work of Zhao {\em et
  al.},\cite{indirect} the additional lowest excitations for
the BL MoS$_2$, WS$_2$ and WSe$_2$ are also reported and claimed to
be the indirect excitation in the $\Gamma$ valley, which is in contrast to the
understanding in the BL TMD
heterostructures.\cite{long_live,inter-layer}
Furthermore, the behavior of the PL depolarization dynamics
for the BL WS$_2$ is revealed to be very different from the ML situation.\cite{Cui}
Zhu {\em et al.}
has observed that with the same experimental conditions, the steady-state PL
polarization for the excitation 1.99 eV (so-called A exciton) is nearly 100$\%$
in the BL WS$_2$,
which is anomalously larger than the one measured in the ML WS$_2$ (less than
$40\%$).\cite{Cui} However, based on the spin-layer locking
picture,\cite{Hamiltonian2}
 this PL
depolarization dynamics is very hard to understand according to the previous study in
ML TMDs,\cite{Yu3,Glazov} where the intrinsic electron-hole (e-h) exchange interaction can cause
efficient PL depolarization due to the Maialle-Silva-Sham
   (MSS) mechanism.\cite{Sham1,Sham2,Cui}

In this paper, we present a possible understanding of the  above
 observations that is different from the above
 speculations.\cite{indirect,long_live,inter-layer,Cui,WS1}
In our picture, the speculated indirect exciton, which involves 
a high-order phonon absorption/emission process, is unnecessary.
 In BL TMDs, due to the strong Coulomb
 interaction,\cite{MoS2,WS1,WS2,WS3,MoSe2,WSe2} the
  e-h pair can form not only the IL exciton but also
the CT one. Furthermore, due to the inter-layer hopping of the
 hole, the IL and CT excitons can couple together to
 form the excimer.\cite{excimer1,excimer2} Here, although the dark exciton can also
   contribute to the formation of the excimer state, on one hand, it has negligible influence
   on the excimer energy level; on the other hand, it cannot be excited in the optical
   process. Hence, in the optical
   process, only the bright exciton needs to be
   considered. Furthermore, in the BL WS$_2$, although there exists 
large energy splitting for the valence bands, due to the anisotropy of dielectric
 constant,\cite{dielectric_B,directgap_wang23,dielectric,dielectric_c} 
 the A IL and B CT
 exciton states are nearly degenerate and hence can 
couple together to form the ${\rm A}'$ and ${\rm B}'$ excimer states with the
 energy level calculated to be 1.99 and 2.10~eV, respectively. 
Accordingly, the binding energy for the excimer states
 are calculated to be 40
  meV, showing good agreement with the observed redshift for the A exciton in the
  BL WS$_2$.\cite{Cui,WS1} Moreover, the energy level for the lowest
  and highest excitations
 are calculated to be 1.69 and 2.41 eV, which correspond to the
 A CT and B IL excitons, also showing good agreement with the
 experiment.\cite{indirect,Cui,WS1} Therefore, according to our
 calculation, the understanding of
 the four excitations is different from the speculation in the experiments, with
 the lowest three excitations being the A CT exciton, ${\rm A}'$ excimer
 and ${\rm B}'$ excimer
 rather than the indirect excitation for the $\Gamma$ valley, trion and A exciton.\cite{indirect,Cui,WS1}

We further study the exchange interaction between the two excimer states
 we reveal in the BL WS$_2$, based on
which we perform the investigation of the PL depolarization dynamics by the
kinetic spin Bloch equations (KSBEs) in the pump-probe setup.\cite{broadening1,broadening2}
 We find both the
Coulomb interaction and inter-layer hopping of the hole can contribute to the
exchange interaction in both the intra- and inter-valley situations. These dominant
processes are illustrated in Fig.~\ref{figyw1} for the intra-valley
situation. On one hand, the e-h pair in one IL exciton can virtually recombine
and then generate another IL exciton due to the Coulomb interaction
directly. On the other hand, there exists another higher-order process, in which
the hole in the CT exciton first hops from one layer to another
and then recombines virtually with the electron part to generate the
IL exciton due to the Coulomb interaction. These exciton transition
processes can cause the excimer transition efficiently due to the MSS
mechanism,\cite{Sham1,Sham2,Yu3,Glazov,Cui}
 with the former process is more important than the latter.

 We then perform the investigation of the PL depolarization dynamics by the
 KSBEs in the pump-probe
 setup.\cite{broadening1,broadening2} The calculations based on the KSBEs show that with the absorption of the
$\sigma_{+}$ light, the emergence of the $\sigma_{-}$ light can be
instantaneous,
which is similar to ML TMDs.\cite{CD,many_body,Yu3,Korn2} Furthermore,  
there is always an anomalous residual PL polarization as large 
as $50\%$ exactly, lasting
for extremely long time, which is robust against the initial energy broadening and strength of the
momentum scattering. This indicates that the
PL depolarization time $\tau_s$ can be much longer than the excimer
lifetime $\tau_r$, which is in the order of
picoseconds.\cite{absorption_Marie,Cui,Korn1,Korn2} Accordingly, based on the rate equation,\cite{absorption_Mak,
absorption_Kioseoglou,valley_wang18,absorption_Marie} this provides a reasonable
explanation for the
anomalously large steady-state PL polarization nearly $100\%$ observed in the
experiment of Zhu {\em et al.} in the BL
  WS$_2$.\cite{Cui} We
further reveal that this anomalous {\em steady state} originates
from the specific form of the exchange interaction Hamiltonian. It is
  interesting to see that there exists a density matrix in the steady
state but with residual PL polarization, which can commute with the exchange interaction Hamiltonian and
hence protects the
large residual PL polarization. Moreover, for the system pumped by the elliptically polarized
light, we demonstrate that the residual
  PL polarization is always half of the initial polarization of the elliptically polarized
light.

This paper is organized as follows. In Sec.~{\ref{model}}, we set up the
  model and lay out the formalism. In Sec.~{\ref{Model_A}}, we derive the
  excimer state and calculate the excimer excitation energy. In Sec.~{\ref{Model_B}},
  we derive the excimer exchange interaction. In Sec.~{\ref{Model_C}}, we present the KSBEs and
    perform the calculations for the PL depolarization dynamics in the pump-probe setup.
 We conclude and discuss in Sec.~{\ref{summary}}.

\section{Model and Formalism} 
\label{model}
In this section, following the previous works within the
framework of effective-mass approximation,\cite{exchange,Sham1,dot,Yu3}
 the excimer Hamiltonian for the envelope wavefunction is derived [refer to
Eq.~(\ref{effective_H}) in Appendix~\ref{AA}]. Based on the excimer Hamiltonian,
we then calculate the energy spectra of
the optical excitations (Sec.~{\ref{Model_A}}) and the exchange interaction
between the excimer states (Sec.~{\ref{Model_B}}) in the BL WS$_2$.
\subsection{Excimer state in BL WS$_2$}
\label{Model_A}
In this subsection, we present the optical excitations in the BL WS$_2$.
 Due to the
strong Coulomb interaction, the e-h pair forms the IL and CT excitons.
Furthermore, due to the efficient inter-layer hopping of the hole, the excimer state can
be formed from the superposition of the IL and CT exciton states.
 Here, although the dark exciton can also
   contribute to the formation of the excimer state, on one hand, it has negligible influence
   on the excimer energy level; on the other hand, it cannot be excited in the optical
   process. Hence, in the optical
   process, only the bright exciton is 
   considered and the corresponding excimer state is referred to as bright excimer.
 Specifically, for the optical excitation experiment with
one photon process, we focus on the excimer ground state ($1s$-state)
$|mn;mn';{\bf P}=0\rangle$ (${\bf P}$ is the center-of-mass momentum),
which is written as 
\begin{eqnarray}
\nonumber
&&\langle {\bf r}_1,{\bf r}_2|mn;mn';{\bf P}=0\rangle=A_{1s}^{mn}f_{1s}^{mn}({\bf
    r}_1-{\bf r}_2)\Psi_{m}({\bf r}_1)\\
&&\mbox{}\times\tilde{\Psi}_{n}({\bf r}_2)+A_{1s}^{mn'}f_{1s}^{mn'}({\bf r}_1-{\bf r}_2)\Psi_{m}({\bf
  r}_1)\tilde{\Psi}_{n'}({\bf r}_2)
\label{excimer}
\end{eqnarray}
in the coordinate representation, 
with the first (second) term at the right hand side of Eq.~({\ref{excimer}}) describing
the IL (CT) exciton wavefunction. Here,
${\bf r}_1$ and ${\bf r}_2$ are the electron and hole coordinates. $m$ in the conduction band and $n$
($n'$) in the valence band
denote the indices including the layer, valley and spin degrees of the
electron; $n$ and $n'$ are limited in the same valley and different layers
  with the spin degrees being the same as the one in $m$.
 $A_{1s}^{mn}$ and $A_{1s}^{mn'}$ represent the amplitudes of the
IL and CT exciton states in the excimer state.
 $f_{1s}^{mn(n')}({\bf r}_1-{\bf r}_2)$ is the two-dimensional hydrogen-like exciton state of
the e-h pair for the electron and hole sitting in the $m$- and $n(n')$-band, and 
when the center-of-mass momentum $\bf P$=0, 
it is written
as, 
\begin{equation} 
f^{mn(n')}_{1s}({\bf r}_1-{\bf r}_2)=\sqrt{8/{\pi
      a_{B\Box}^2}}\exp({-2|{\bf r}_1-{\bf r}_2|/a_{B\Box}}),
\end{equation}
with $a_{B\Box}$ being the Bohr radii for the exciton, which are different for the
IL and CT excitons represented by $a_{B\parallel}$ and $a_{B\perp}$, respectively, due to
 the anisotropy of the dielectric
 constant.\cite{dielectric_B,directgap_wang23,dielectric,dielectric_c} In
   the following, the hydrogen-like exciton state for the IL and CT excitons are
 further explicitly represented by $f^{\rm IL}_{1s}({\bf r}_1-{\bf r}_2)$ and $f^{\rm CT}_{1s}({\bf r}_1-{\bf r}_2)$. 
 $\Psi_m({\bf r})$ [$\tilde{\Psi}_{n(n')}({\bf
  r})$] is the band-edge wavefunction for the electron (hole).\cite{exchange,Sham1,exchange,dot,Yu3}

From Eq.~(\ref{effective_H}), the amplitudes $A_{1s}^{mn(n')}$ satisfy the equation
\begin{equation}
(E_{m}-E_{n}+E_{1s}^{mn})A_{1s}^{mn}+\sum_{n'}\tilde{T}_{nn'}A_{1s}^{mn'}=EA_{1s}^{mn},
\label{eigen}
\end{equation}
with $E_m$ and $E_n$ being the band-edge energies for the $m$- and $n$-band,
respectively; $E_{1s}^{mn}$ representing the exciton binding energy for the $m$-band
electron and $n$-band hole, which is further denoted by $E_{b,\parallel}$ and
  $E_{b,\perp}$ for the IL and CT excitons; $\tilde{T}_{nn'}$ standing for the
effective hopping energy for the hole between $n$- and $n'$-band, which only
exists for the hole in the same valley with the same spin between different layers. This
 effective hopping energy $\tilde{T}_{nn'}$ for the hole is determined by the overlap of the
 IL and CT hydrogen-like exciton wavefunctions, written as
\begin{eqnarray}
\nonumber
&&\tilde{T}_{nn'}\equiv t_{\perp}^*=t_{\perp}\int d{\bf r}_1d{\bf
  r}_2f_{1s}^{mn}({\bf r}_1-{\bf r}_2)f_{1s}^{mn'}({\bf r}_1-{\bf r}_2)\\
&&\mbox{}=4t_{\perp}a_{B\parallel}a_{B\perp}/(a_{B\parallel}+a_{B\perp})^2,
\label{e_hopping}
\end{eqnarray}
with $t_{\perp}$ being the inter-layer hoping energy for the
hole\cite{Hamiltonian1,Hamiltonian2}. 
Finally, the eigen-equations [Eq.~(\ref{eigen})] for the
 amplitudes of the IL ($A_{1s}^{mn}$) and CT ($A_{1s}^{mn'}$)
 excitons in the excimer state are written in the matrix
form as 
\begin{equation}
\left(\begin{array}{cc}
E_1-E & \tilde{T}_{nn'} \\
\tilde{T}_{n'n} & E_2-E
\end{array}\right)
\left(\begin{array}{c}
A_{1s}^{mn} \\
A_{1s}^{mn'}
\end{array}\right)=0.
\label{equation}
\end{equation}
Here, $E_1=E_{m}-E_{n}+E_{b,\parallel}$ and $E_2=E_{m}-E_{n'}+E_{b,\perp}$
stand for the energy levels for the IL and CT excitons, respectively.
Specifically, there are 16 configurations for the bright exciton states in BL TMDs, in which the 4 degenerate A
  or B IL (CT) excitons are distinguished by the valley and layer degrees of freedom.
 Therefore, actually there are only four kinds of bright exciton states and hence
 two kinds of bright excimer
 states needed to be considered. One excimer state is composed
   of the A IL and B CT excitons and the other one is composed
   of the B IL and A CT excitons.
 Obviously, although there exists large energy splitting for the valence bands,
 by considering different binding energies $E_{1s}^{mn}$ and
 $E_{1s}^{mn'}$ for the IL and
 CT excitons due to the
anisotropy of the dielectric constant,\cite{dielectric_B,directgap_wang23,dielectric,dielectric_c} the energy
levels $E_1$ and $E_2$ for the IL and CT excitons 
 can be close to each other.  

From Eq.~(\ref{equation}), by assuming
$E_1<E_2$, the eigenvalues, which denote the excimer excitation energies, are
written as
\begin{equation}
\left\{\begin{array}{c}
E_{1'}=\frac{\displaystyle E_1+E_2}{\displaystyle 2}-\frac{\displaystyle
  1}{\displaystyle 2}\sqrt{(E_1-E_2)^2+4|t_{\perp}^*|^2}\\
E_{2'}=\frac{\displaystyle E_1+E_2}{\displaystyle 2}+\frac{\displaystyle
  1}{\displaystyle 2}\sqrt{(E_1-E_2)^2+4|t_{\perp}^*|^2}
\end{array}\right..
\label{eigen_values}
\end{equation}
Hence the IL and CT excitons can
couple together and the excimer state is formed when $|E_1-E_2|\ll 2t_{\perp}^*$.  Accordingly,
the amplitudes $A_{i'}^{\rm IL(CT)}$ of the IL (CT) exciton state in the excimer
state with excitation energy $E_{i'}$ are expressed as
\begin{equation}
\left\{\begin{array}{c}
A_{i'}^{\rm IL}=\frac{\displaystyle t^{*}_{\perp}}{\displaystyle \sqrt{(t^{*}_{\perp})^2+(E_i-E_{i'})^2}} \\
A_{i'}^{\rm CT}=\frac{\displaystyle E_{i'}-E_i}{\sqrt{\displaystyle (t^{*}_{\perp})^2+(E_i-E_{i'})^2}}
\end{array}\right..
\label{amplitude1}
\end{equation}

In the following, we first list 
the material
parameters used in the calculation of excimer state, shown in
Table~\ref{parameter1}.\cite{Hamiltonian1,Hamiltonian2,dielectric,MoS2,WS1,WS2,WS3,MoSe2,WSe2}
In Table~\ref{parameter1}, the dielectric constants for the IL and CT
exciton are denoted by $\kappa_{\parallel}$ and
$\kappa=\sqrt{\kappa_{\parallel}\kappa_{\perp}}$, respectively,\cite{dielectric_B,directgap_wang23,dielectric,dielectric_c}
where $\kappa_{\parallel}$ and $\kappa_{\perp}$ represent the dielectric constants
parallel and perpendicular to the layer, respectively. 
 Accordingly, by using the binding energy of the IL exciton
 $E_{b,\parallel}$ directly measured from the experiments in ML
TMDs,\cite{MoS2,WS1,WS2,WS3,MoSe2,WSe2} the binding energy of the CT
 exciton $E_{b,\perp}$ and the exciton Bohr radius $a_{B\parallel}$
($a_{B\perp}$)
 for the IL (CT)
exciton are calculated, as shown in
Table~\ref{parameter1}.\cite{directgap_wang23} Furthermore, the effective
hopping energy for the hole $t_{\perp}^*$ is determined according to Eq.~(\ref{e_hopping}) with
$t_{\perp}$ known.\cite{Hamiltonian1, Hamiltonian2} The experimentally
 measured energy gap between the lowest conduction band and highest
valence band,\cite{MoS2,WS1,WS2,WS3,MoSe2,WSe2} and energy splitting
 for the valence bands
 are also listed in Table~\ref{parameter1}.\cite{Hamiltonian1,Hamiltonian2,WS1,WS2,WS3,WSe2} 

\begin{table}[htb]
  \caption{Material parameters for the calculation of the excimer state and excitation
energy.}
  \label{parameter1} 
  \begin{tabular}{l l l l l}
    \hline
    \hline
    &\;\;\;\;{MoS$_2$}\;\;\;&\;\;\;\;{WS$_2$}\;\;\;&\;\;\;\;\;{MoSe$_2$}\;\;\;&\;\;\;\;\;{WSe$_2$}\\  
    \hline 
    $\kappa_{\parallel}$&\;\;\;\;\;$4.8$&\;\;\;\;\;$4.4$&\;\;\;\;\;\;$6.9$&\;\;\;\;\;\;$4.5$\\
    $\kappa_{\perp}$&\;\;\;\;\;$3.0$&\;\;\;\;\;$2.9$&\;\;\;\;\;\;$3.8$&\;\;\;\;\;\;$2.9$\\
    $\kappa$&\;\;\;\;\;$3.8$&\;\;\;\;\;$3.6$&\;\;\;\;\;\;$5.1$&\;\;\;\;\;\;$3.6$\\
    $E_{b,\parallel}$ (meV)&\;\;\;\;\;$570^a$&\;\;\;\;\;$700^{b}$&\;\;\;\;\;\;$550^c$&\;\;\;\;\;\;$600^{d}$\\ 
    $E_{b,\perp}$ (meV)&\;\;\;\;\;909&\;\;\;\;\;1046&\;\;\;\;\;\;1007&\;\;\;\;\;\;938\\
    $a_{B\parallel}$ (${\rm \AA}$)&\;\;\;\;\;$10.5$&\;\;\;\;\;$9.4$&\;\;\;\;\;\;$13.9$&\;\;\;\;\;\;10.7 \\
    $a_{B\perp}$ (${\rm \AA}$)&\;\;\;\;\;$8.3$&\;\;\;\;\;$7.7$&\;\;\;\;\;\;$5.6$&\;\;\;\;\;\;8.5 \\
    2$t_{\perp}$ (meV)&\;\;\;\;\;$86$&\;\;\;\;\;$109$&\;\;\;\;\;\;$106$&\;\;\;\;\;\;134\\
    2$t_{\perp}^*$ (meV)&\;\;\;\;\;$79$&\;\;\;\;\;$102$&\;\;\;\;\;\;$101$&\;\;\;\;\;\;124\\
    $E_g$
    (eV)&\;\;\;\;\;$2.5^a$&\;\;\;\;\;$2.73^b$&\;\;\;\;\;\;$2.18^c$&\;\;\;\;\;\;$2.4^d$\\
    2$\lambda_v$ (meV)&\;\;\;\;\;$147$&\;\;\;\;\;$380^b$&\;\;\;\;\;\;$182$&\;\;\;\;\;\;$420^{d}$ \\
    \hline
    \hline
\end{tabular}\\
$^a$ Ref.~\onlinecite{MoS2}. \quad$^b$
 Refs.~\onlinecite{WS1,WS2,WS3}. \quad$^c$ Ref.~\onlinecite{MoSe2}.
 \quad $^d$ Ref.~\onlinecite{WSe2}. 
\end{table}

Based on the material parameters in Table~\ref{parameter1}, for the BL WS$_2$, 
the energy levels $E_{1}$ and $E_2$ for
the IL and CT excitons are further calculated, shown in
  Table~\ref{parameter2}. 
In Table~\ref{parameter2}, the two kinds of excimer states in BL TMDs 
are further represented by $|{\rm IL}_{A}; {\rm CT}_{B}\rangle$ and $|{\rm
  IL}_{B}; {\rm CT}_{A}\rangle$, respectively.
We then calculate the optical excitation energies $E_{1'}$ and $E_{2'}$ for the
excimer from
    Eq.~(\ref{eigen_values}), 
 and the corresponding
  amplitudes ${A}^{\rm IL(CT)}_{1'}$ and ${A}^{\rm IL(CT)}_{2'}$ of
the IL (CT) exciton in the excimer state from Eq.~(\ref{amplitude1}),
as shown in Table~\ref{parameter2}.

From the results in Table~\ref{parameter2}, it can be seen that there are four optical excitations
in the BL WS$_2$,
 whose energy levels are calculated to be 1.69, 1.99, 2.10 and 2.41~eV,
respectively. On one hand, for the lowest (highest) energy level 1.69 (2.41)~eV,
it can be seen from the amplitude $A_{2'}^{\rm CT}\approx 1$
  ($A_{1'}^{\rm IL}\approx 1$) that the excitation state is
actually the A CT (B IL) exciton state. On the other hand,
 the energy levels $E_1=2.03$ eV for the A IL exciton and $E_2=2.06$ eV for the B CT
exciton are very close ($|E_1-E_2|\ll 2t^*_{\perp}$), and hence the ${\rm A}'$
and ${\rm B}'$
excimer states corresponding to $E_1'=1.99$ and $E_2'=2.10$~eV form due to the
efficient inter-layer hopping of the hole. Accordingly, the binding energy for
the ${\rm A}'$ excimer state is calculated to be
$|E_1-E_1'|=40$~meV. Our calculated results show good agreement
with the recent experiments in BL WS$_2$.\cite{indirect,Cui,WS1}

In the experiment of Zhu {\em et al.} for the BL WS$_2$,\cite{Cui} it
has been observed that there are four resonant excitations with excitation
energies approximately being 1.68, 1.93, 1.99 and
2.37~eV, in good agreement with our calculation with 1.69, 1.99, 2.10 and
 2.41~eV.\cite{indirect}
However, in the experiments,\cite{indirect,Cui,WS1}
 these four excitations have been speculated to be the indirect exciton for the $\Gamma$ valley,
trion, A exciton
and B exciton excitations, respectively. According to our theory, these
four excitations are the A CT exciton, ${\rm A}'$ excimer, ${\rm B}'$ excimer and B IL
exciton, in consistence with the understanding in the BL
 TMD heterostructures.\cite{long_live,inter-layer}
 Furthermore, compared
to the ML WS$_2$, the obvious redshift for the A exciton about 30 $\sim$ 
50~meV is observed in the BL WS$_2$ in these experiments,\cite{Cui, WS1}
which also confirms our calculated binding energy $40$~meV for the ${\rm A'}$
excimer.

 Finally, we address that the inter-layer hopping energy for the hole can be tuned by
variation of the inter-layer distance,\cite{excimer2} which can be realized in high pressure
experiment.\cite{excimer1,pressure} Hence, the energy levels of the ${\rm A'}$
and ${\rm B'}$ excimer states
can be
tuned by means of the pressure in BL WS$_2$ and other BL TMDs, which can be
observed in the experiment directly.\cite{excimer1,pressure}

\begin{center}
\begin{table}[htb]
  \caption{Energy levels and
    the amplitudes for the IL and
    CT excitons for the excimer state in the BL WS$_2$. $E_{1}$ and $E_2$ are the
 energy levels for the corresponding IL and CT excitons in the
   excimer states; $E_{1'}$ and $E_{2'}$
 are the optical excitation energies for the excimer with ${A}^{\rm IL(CT)}_{1'}$ and
   ${A}^{\rm IL(CT)}_{2'}$ being the corresponding
  amplitudes of the IL (CT) exciton.}
  \label{parameter2}
  \begin{tabular}{l l l l l}
    \hline
    \hline
    &\;\;\;\;{$|{\rm IL}_A;{\rm CT}_B\rangle$}\;\;\;&\;\;\;\;{$|{\rm IL}_B;{\rm CT}_A\rangle$}\\  
    \hline
    $E_1$ (eV)&\;\;\;\;\;\;\;\;2.41 &\;\;\;\;\;\;\;\;2.03  \\  
    $E_2$ (eV)&\;\;\;\;\;\;\;\;1.69 &\;\;\;\;\;\;\;\;2.06  \\
    $E_{1'}$ (eV)&\;\;\;\;\;\;\;\;2.41 &\;\;\;\;\;\;\;\;1.99  \\
    $E_{2'}$ (eV)&\;\;\;\;\;\;\;\;1.69 &\;\;\;\;\;\;\;\;2.10  \\
    \hline
    $A^{\rm IL}_{1'}$&\;\;\;\;\;\;\;\;1.0 &\;\;\;\;\;\;\;\;0.79  \\
    $A^{\rm CT}_{1'}$&\;\;\;\;\;\;\;\;0.0 &\;\;\;\;\;\;\;\;-0.62  \\
    $A^{\rm IL}_{2'}$&\;\;\;\;\;\;\;\;0.0 &\;\;\;\;\;\;\;\;0.62  \\
    $A^{\rm CT}_{2'}$&\;\;\;\;\;\;\;\;1.0 &\;\;\;\;\;\;\;\;0.79  \\
    \hline
    \hline
\end{tabular}
\end{table}
\end{center}

\subsection{Exchange interaction between excimer states}
\label{Model_B}
We further show the exchange interaction between the excimer states 
in BL TMDs (refer to Appendix~\ref{AA}).\cite{exchange,Sham1,exchange,dot,Yu3}
 With the IL and CT exciton states with center-of-mass ${\bf P}$ expressed as
$|mn;{\bf P}\rangle$ and $|mn';{\bf P}\rangle$ in
Eq.~(\ref{excimer}), for simplicity, the excimer state is further
represented as
\begin{equation}
|mn;mn';{\bf P}\rangle=A_{1s}^{mn}|mn;{\bf P}\rangle+A_{1s}^{mn'}|mn';{\bf P}\rangle.
\end{equation}
Accordingly, the exchange interaction between the two excimer states
$|m_1n_1;m_1n_1';{\bf P}\rangle$
 and $|m_2n_2;m_2n_2';{\bf P}'\rangle$ can be
obtained from the exchange interaction between the exciton states,
which only exists between the {\em bright} ones,\cite{Yu3,Glazov}
\begin{eqnarray}
\nonumber
&&\langle m_2n_2;m_2n_2';{\bf P}'|H^{ex}|m_1n_1;m_1n_1';{\bf P}\rangle\\
\nonumber
&&\mbox{}=(A_{1s}^{m_2n_2})^*A_{1s}^{m_1n_1}\langle m_2n_2;{\bf
    P}'|H^{ex}|m_1n_1;{\bf P}\rangle\\
\nonumber
&&\mbox{}+(A_{1s}^{m_2n_2})^*A_{1s}^{m_1n_1'}\langle m_2n_2;{\bf
    P}'|H^{ex}|m_1n_1';{\bf P}\rangle\\
\nonumber
&&\mbox{}+(A_{1s}^{m_2n_2'})^*A_{1s}^{m_1n_1}\langle m_2n_2';{\bf
    P}'|H^{ex}|m_1n_1;{\bf P}\rangle\\
&&\mbox{}+(A_{1s}^{m_2n_2'})^*A_{1s}^{m_1n_1'}\langle m_2n_2';{\bf
    P}'|H^{ex}|m_1n_1';{\bf P}\rangle.
\label{excimer_exchange}
\end{eqnarray}
At the right hand side of Eq.~(\ref{excimer_exchange}), the first (last) term describes the
exchange interaction between the two IL (CT) exciton
states; whereas
the second and third terms show the exchange interaction between the IL
and CT exciton states. These exchange interactions between the
bright 
exciton states include both the long-rang (L-R) and short-range (S-R)
parts, with the latter one usually being one order of magnitude smaller than the former in
semiconductors.\cite{GaAs} For the L-R part in Eq.~(\ref{excimer_exchange}), as
shown later, the first term
is one order of magnitude larger than the second
and third terms; whereas the second
and third terms are one order of magnitude larger than the last one. For the S-R part, the
exchange interaction between the IL exciton states (the first term) is in the same order
as the L-R one between the IL and CT excitons (the second and third terms). Here, we
only show the explicit form for the exchange interaction in the same order as the L-R one between the
IL and CT excitons, which are dominant in the exchange interaction
between the 
excimers (the second and third terms).

With the initial and final exciton states being the IL bright exciton
states, the exchange interaction describes the virtual recombination of the e-h
pair in one IL bright exciton
state and then generation of another IL bright one due to the Coulomb
interaction directly (as shown in Fig.~\ref{figyw1}).
 The L-R (S-R) exchange interaction is written as [given in
 Eq.~(\ref{SR}) in Appendix~\ref{AA}]
\begin{equation}
H^{L(1)}_{m'n'\atop
  mn}=\frac{e^2}{2\varepsilon_0\kappa_{\parallel}|{\bf P}|}\delta_{{\bf
  P},{\bf P}'}\big[f_{1s}^{\rm IL}(0)\big]^*f_{1s}^{\rm IL}(0)Q^{(1)}_{m'n'\atop
  mn}({\bf P}),
\label{long_range}
\end{equation}
where
\begin{eqnarray}
\nonumber
&&Q^{(1)}_{m'n'\atop
  mn}({\bf
  P})=\frac{\hbar^2}{2m_0^2}\Big[\frac{1}{(E_m-E_n)^2}+\frac{1}{(E_{m'}-E_{n'})^2}\Big]\\
&&\mbox{}\times({\bf
P}\cdot {\bgreek \pi_{m'\Theta n'}})({\bf
P}\cdot {\bgreek \pi_{\Theta n m}}).
\end{eqnarray}
Here, $m_0$ is the free electron mass;  $\varepsilon_0$ stands for the vacuum
permittivity; ${\bgreek \pi_{m' \Theta
    n'}}$ and ${\bgreek \pi_{\Theta n m}}$ come from the ${\bf k}\cdot{\bf p}$ matrix
elements in the Hamiltonian [Eq.~(\ref{Hamiltonian})] 
with $\Theta$ being the time reversal operator (refer to Appendix~\ref{AA}).

With the IL bright exciton states shown in the order $\mid\uparrow_{\rm Kc}^u,\uparrow_{\rm Kv}^u\rangle$, $\mid\downarrow_{\rm
  Kc}^u,\downarrow_{\rm Kv}^u\rangle$, $\mid\uparrow_{\rm Kc}^l,\uparrow_{\rm Kv}^l\rangle$,
$\mid\downarrow_{\rm Kc}^l,\downarrow_{\rm Kv}^l\rangle$, $\mid\uparrow_{\rm K'c}^u,\uparrow_{\rm K'v}^u\rangle$, $\mid\downarrow_{\rm
  K'c}^u,\downarrow_{\rm K'v}^u\rangle$, $\mid\uparrow_{\rm K'c}^l,\uparrow_{\rm
  K'v}^l\rangle$ and $\mid\downarrow_{\rm K'c}^l,\downarrow_{\rm K'v}^l\rangle$, according to
Eq.~(\ref{long_range}), the exchange
  interaction matrix between the above IL bright exciton states
 $|m_1n_1;{\bf P}\rangle$ and $|m_2n_2;{\bf P}'\rangle$ reads  
\begin{widetext}
\begin{equation}
H^{(1)}_{\rm ex}=\frac{C\delta_{\bf P, P'}}{|{\bf P}|}\left(\begin{array}{cccccccc}
\alpha_1|{\bf P}|^2 & \beta|{\bf P}|^2 & \beta P_{+}^2 & \alpha_1 P_{+}^2
 & -\beta P_{+}^2 & -\alpha_1P_{+}^2 & -\alpha_1|{\bf P}|^2 & -\beta|{\bf P}|^2\\
 & \alpha_2|{\bf P}|^2 & \alpha_2P_{+}^2  & \beta P_{+}^2 &
 -\alpha_2 P_{+}^2 & -\beta P_{+}^2 &-\beta|{\bf P}|^2  &-\alpha_2|{\bf P}|^2 \\
 &   & \alpha_2|{\bf P}|^2  & \beta |{\bf P}|^2 &
-\alpha_2|{\bf P}|^2  & -\beta|{\bf P}|^2 & -\beta P_{-}^2 & -\alpha_2 P_{-}^2\\
 &   &   & \alpha_1|{\bf P}|^2 &-\beta|{\bf P}|^2  & -\alpha_1|{\bf P}|^2 &-\alpha_1 P_{-}^2  &-\beta P_{-}^2 \\
 &   &   &  &\alpha_2|{\bf P}|^2  & \beta|{\bf P}|^2 & \beta P_{-}^2 & \alpha_2 P_{-}^2\\
 &   &   &  &  & \alpha_1|{\bf P}|^2 & \alpha_1P_{-}^2 & \beta P_{-}^2\\
 &   &   &  &  &  & \alpha_1|{\bf P}|^2 & \beta |{\bf P}|^2\\
 &   &   &  &  &  &  & \alpha_2|{\bf P}|^2
\end{array}\right).
\label{Coulomb}
\end{equation}
\end{widetext}
Here, $C=e^2/(2
    \varepsilon_0\kappa_{\parallel})\big|f_{1s}^{\rm IL}(0)\big|^2$;
 $P_{\pm}=P_x\pm iP_y$; the material parameters $\alpha_1=a^2t^2/(\Delta-\lambda_c+\lambda_v)^2$,
$\alpha_2=a^2t^2/(\Delta+\lambda_c-\lambda_v)^2$ 
and
$\beta=\frac{1}{2}\big[\frac{a^2t^2}{(\Delta+\lambda_c-\lambda_v)^2}+\frac{a^2t^2}{(\Delta-\lambda_c+\lambda_v)^2}\big]$
are calculated according to the parameters in
Ref.~\onlinecite{Hamiltonian1} with $\lambda_c$ representing the splitting of
the conduction band,
 shown in Table~\ref{parameter_ex} for the BL MoS$_2$,
WS$_2$, MoSe$_2$ and WSe$_2$, respectively.    

With the initial and final exciton states being the CT and
IL exciton states, the exchange interaction describes the process that 
the hole in the CT exciton first hops from
one layer to another and then recombines virtually with
the electron part to generate the IL exciton
due to the Coulomb interaction.
 Hence, this process [Eq.~(\ref{broken})] for the L-R exchange interaction is in
 the order $t_{\perp}^*/E_g$ times
of the former L-R one [Eq.~(\ref{long_range})]. This L-R exchange interaction is expressed as
\begin{equation}
\nonumber
H^{L(2)}_{m'n'\atop
  mn}=\frac{e^2}{2\varepsilon_0\kappa_{\parallel}|{\bf P}|}\delta_{{\bf
  P},{\bf P}'}\big[f_{1s}^{\rm IL}(0)\big]^*f_{1s}^{\rm CT}(0)Q^{(2)}_{m'n'\atop
  mn}({\bf P}),
\label{broken}
\end{equation}
where
\begin{eqnarray}
\nonumber
&&Q^{(2)}_{m'n'\atop
  mn}({\bf
  P})=\frac{\hbar^2}{2m_0^2}\frac{({\bf
P}\cdot {\bgreek \pi_{m'\Theta n'}})({\bf
P}\cdot {\bgreek \pi_{\Theta l'' m}})\tilde{T}_{\Theta n \Theta
l''}}{(E_m-E_{l''})(E_m-E_n)}\\
&&\mbox{}\times\Big(\frac{1}{E_m-E_{l''}}+\frac{1}{E_{m}-E_{n}}\Big).
\end{eqnarray}

With the CT exciton states being represented by $\mid\uparrow_{\rm Kc}^u,\uparrow_{\rm Kv}^l\rangle$, $\mid\downarrow_{\rm
  Kc}^u,\downarrow_{\rm Kv}^l\rangle$, $\mid\uparrow_{\rm Kc}^l,\uparrow_{\rm Kv}^u\rangle$,
$\mid\downarrow_{\rm Kc}^l,\downarrow_{\rm Kv}^u\rangle$, $\mid\uparrow_{\rm K'c}^u,\uparrow_{\rm K'v}^l\rangle$, $\mid\downarrow_{\rm
  K'c}^u,\downarrow_{\rm K'v}^l\rangle$, $\mid\uparrow_{\rm K'c}^l,\uparrow_{\rm
  K'v}^u\rangle$ and $\mid\downarrow_{\rm K'c}^l \downarrow_{\rm K'v}^u\rangle$,
according to Eq.~(\ref{broken}), the exchange interaction matrix
between the above CT bright exciton and IL bright exciton states $|mn;{\bf P}\rangle$
and $|mn';{\bf P'}\rangle$ reads
\begin{widetext}
\begin{equation}
H^{(2)}_{\rm ex}=\frac{C'\delta_{\bf P, P'}}{|{\bf P}|}\left(\begin{array}{cccccccc}
\tilde{\alpha}(1)|{\bf P}|^2 & \tilde{\beta}(1)|{\bf P}|^2 & \tilde{\beta}(-1)P_{+}^2 & \tilde{\alpha}(-1)P_{+}^2 & -\tilde{\beta}(1)P_{+}^2 & -\tilde{\alpha}(1)P_{+}^2 & -\tilde{\alpha}(-1)|{\bf P}|^2 & -\tilde{\beta}(-1)|{\bf P}|^2\\
 & \tilde{\beta}(1)|{\bf P}|^2 & \tilde{\beta}(-1)P_{+}^2  & \tilde{\alpha}(-1)P_{+}^2 & -\tilde{\beta}(1)P_{+}^2 & -\tilde{\alpha}(1)P_{+}^2 & -\tilde{\alpha}(-1)|{\bf P}|^2 & -\tilde{\beta}(-1)|{\bf P}|^2\\
 &   & \tilde{\beta}(-1)|{\bf P}|^2  & \tilde{\alpha}(-1)|{\bf P}|^2 & -\tilde{\beta}(1)|{\bf P}|^2 & -\tilde{\alpha}(1)|{\bf P}|^2 & -\tilde{\alpha}(-1)P_{-}^2 & -\tilde{\beta}(-1)P_{-}^2\\
 &   &   & \tilde{\alpha}(-1)|{\bf P}|^2 & -\tilde{\beta}(1)|{\bf P}|^2 & -\tilde{\alpha}(1)|{\bf P}|^2 & -\tilde{\alpha}(-1)P_{-}^2 & -\tilde{\beta}(-1)P_{-}^2\\
 &   &   &  & \tilde{\beta}(1)|{\bf P}|^2 & \tilde{\alpha}(1)|{\bf P}|^2 & \tilde{\alpha}(-1)P_{-}^2 & \tilde{\beta}(-1)P_{-}^2\\
 &   &   &  &  & \tilde{\alpha}(1)|{\bf P}|^2 & \tilde{\alpha}(-1)P_{-}^2 & \tilde{\beta}(-1)P_{-}^2\\
 &   &   &  &  &  & \tilde{\alpha}(-1)|{\bf P}|^2 & \tilde{\beta}(-1)|{\bf P}|^2\\
 &   &   &  &  &  &   & \tilde{\beta}(-1)|{\bf P}|^2
\end{array}\right).
\label{hopping}
\end{equation}
\end{widetext}
Here, $C'=e^2/(2
    \varepsilon_0\kappa_{\parallel})\big[f_{1s}^{\rm
        IL}(0)\big]^*f_{1s}^{\rm CT}(0)$.
 The parameters
\begin{equation}
\tilde{\alpha}(\tau)=\frac{a^2t^2\tau_{\perp}^*(\Delta-\lambda_c+\tau
  Ed/2)}{(\Delta-\lambda_c+\lambda_v)^2(\Delta-\lambda_c-\lambda_v+\tau Ed)^2}
\end{equation}
and
\begin{equation}
\tilde{\beta}(\tau)=\frac{a^2t^2\tau_{\perp}^*(\Delta+\lambda_c+\tau
  Ed/2)}{(\Delta+\lambda_c-\lambda_v)^2(\Delta+\lambda_c+\lambda_v+\tau Ed)^2},
\end{equation}
with $E$ and
$d$
 being the magnitude of the electric field
and the interlayer distance, respectively. Specifically, one observes that
  the form of Eq.~(\ref{hopping}) is very similar to the one of
  Eq.~(\ref{Coulomb}), with the magnitude of the former [Eq.~(\ref{hopping})] being one order smaller
  than the latter [Eq.~(\ref{Coulomb})].
 $\tilde{\alpha}(\tau)$ and $\tilde{\beta}(\tau)$ are calculated with
the material parameters taken from Ref.~{\onlinecite{Hamiltonian1}} for $E=0$, 
 shown in Table~\ref{parameter_ex} for the BL MoS$_2$,
WS$_2$, MoSe$_2$ and WSe$_2$, respectively.

\begin{table}[htb]
  \caption{Material parameters $\alpha_1$, $\alpha_2$, $\beta$,
    $\tilde{\alpha}(\pm 1)$ and $\tilde{\beta}(\pm 1)$ for the BL MoS$_2$,
WS$_2$, MoSe$_2$ and WSe$_2$ with the unit being ${\rm \AA}^2$.}
  \label{parameter_ex} 
  \begin{tabular}{l l l l l l}
    \hline
    \hline
    &\;\;\;\;\;\;{$\alpha_1$}\;\;\;&\;\;\;\;\;{$\alpha_2$}\;\;\;&\;\;\;\;\;\;{$\beta$}&\;\;\;\;\;{$\tilde{\alpha}(\pm
      1)$}&\;\;\;\;\;{$\tilde{\beta}(\pm
      1)$}\\  
    \hline
    MoS$_2$&\;\;\;\;\;$4.51$&\;\;\;\;$3.82$&\;\;\;\;$4.16$&\;\;\;\;\;\;$0.09$&\;\;\;\;\;\;$0.09$\\
    WS$_2$&\;\;\;\;\;$6.54$&\;\;\;\;$4.43$&\;\;\;\;$5.48$&\;\;\;\;\;\;$0.14$&\;\;\;\;\;\;$0.14$\\
    MoSe$_2$&\;\;\;\;\;$4.65$&\;\;\;\;$3.67$&\;\;\;\;$4.16$&\;\;\;\;\;\;$0.14$&\;\;\;\;\;\;$0.14$\\
    WSe$_2$&\;\;\;\;\;$7.39$&\;\;\;\;$4.49$&\;\;\;\;$5.94$&\;\;\;\;\;\;$0.21$&\;\;\;\;\;\;$0.21$\\
    \hline
    \hline
\end{tabular}
\end{table}  

From Eqs.~(\ref{Coulomb}) and (\ref{hopping}), both the intra-
  and inter-valley exchange interactions can cause the bright excimer transition
  due to the
MSS mechanism.\cite{Sham1,Sham2} However, if the energy levels for the two
excimer states have large splitting, the excimer transition is nearly forbidden due to the
  detuning effect.\cite{Haug} Hence, in the BL WS$_2$,
 by considering the large energy splitting about 100 meV for the ${\rm
      A}'$ and ${\rm B}'$ excimer states, we only need to consider the
  transition between the degenerate
  excimer states.\cite{Cui}

\section{PL depolarization due to e-h exchange interaction} 
\label{Model_C}
In this section, we focus on the PL depolarization due to the e-h exchange
interaction based on the KSBEs in the BL WS$_2$. We first present the model and
then study the PL depolarization dynamics in the pump-probe setup.
\subsubsection{Model and KSBEs}
We focus on
the four degenerate ${\rm A}'$ bright excimer states ($E_1'=1.99$ eV) according to
the experiment condition in the work of Zhu {\em et al.},\cite{Cui} which are represented as 
$\mid\downarrow_{\rm Kc}^{u},\downarrow_{\rm Kv}^{u};\downarrow_{\rm
  Kc}^{u},\downarrow_{\rm Kv}^{l};\bf P\rangle$,
 $\mid\uparrow_{\rm Kc}^{l},\uparrow_{\rm Kv}^{l};\uparrow_{\rm
   Kc}^{l},\uparrow_{\rm Kv}^{u};\bf P\rangle$,
 $\mid\uparrow_{\rm K'c}^{u},\uparrow_{\rm K'v}^{u};\uparrow_{\rm
   K'c}^{u},\uparrow_{\rm K'v}^{l};\bf P\rangle$
 and $\mid\downarrow_{\rm K'c}^{l},\downarrow_{\rm K'v}^{l};\downarrow_{\rm
   K'c}^{l},\downarrow_{\rm K'v}^{u};\bf P\rangle$. According to chiral optical valley
selection rule,\cite{Hamiltonian1} the first and fourth (second and third) states are associated
with $\sigma_{+}$ ($\sigma_{-}$) light. From Eqs.~(\ref{Coulomb}) and
(\ref{hopping}), the L-R exchange interaction between the
  four excimer states are written as
\begin{equation}
H_{\rm ex}({\bf P})\approx\frac{C''\delta_{\bf P, P'}}{|{\bf P}|}\left(\begin{array}{cccc}
|{\bf P}|^2 & P_{+}^2 & -P_{+}^2 & -|{\bf P}|^2\\
P_{-}^2 & |{\bf P}|^2 & -|{\bf P}|^2  & -P_{-}^2\\
-P_{-}^2&-|{\bf P}|^2   & |{\bf P}|^2  & P_{-}^2\\
-|{\bf P}|^2 & -P_{+}^2  & P_{+}^2  & |{\bf P}|^2\\
\end{array}\right).
\label{Coulomb2}
\end{equation} 
Here, $C''=\big|A_{1'}^{\rm IL}\big|^2C\alpha_2+2A_{1'}^{\rm
    IL}A_{1'}^{\rm CT}C'\tilde{\beta}(\pm 1)$.

With the exchange interaction Hamiltonian [Eq.~(\ref{Coulomb2})], the PL
depolarization dynamics associated with the ${\rm A}'$ bright excimers
($E_1'=1.99$ eV)
 can be described by the KSBEs:\cite{Sham1,Sham2,broadening1,broadening2}
\begin{equation}
  \partial_t\rho({\bf P},t)=\partial_t\rho({\bf
      P},t)|_{\rm coh}+\partial_t\rho({\bf P},t)|_{\rm  scat}.
\label{ksbe}
\end{equation}
In these equations, $\rho({\bf P},t)$  represent the $4\times4$ density matrices
of the ${\rm A}'$ bright excimers with center-of-mass momentum ${\bf P}$ at time $t$,
 in which the diagonal elements $\rho_{ss}({\bf P},t)$ describe the excimer distribution
 functions and the off-diagonal elements $\rho_{ss'}({\bf P},t)$ with $s\ne s'$ 
represent the coherence between different excimer states. In the
  collinear space, the coherent term is given by  
\begin{equation}
\partial_t\rho({\bf P},t)|_{\rm
   coh}=-\frac{i}{\hbar}\big[H_{\rm ex},\rho({\bf
  P},t)\big],
\end{equation}
where $[\ ,\ ]$ denotes the commutator.
 The scattering term
 $\partial_t\rho({\bf P},t)|_{\rm  scat}$ is written in the elastic
 approximation as\cite{Sham1} 
 \begin{equation}
\partial_t\rho({\bf P},t)|_{\rm
   scat}=\sum_{{\bf P}'}W_{{\bf P}{\bf P}'}\big[\rho({\bf P}',t)-\rho({\bf
  P},t)\big].
\label{scat}
\end{equation}
Here, $W_{{\bf P}{\bf P}'}$ represents the momentum scattering rate.

By solving the KSBEs, one
obtains the evolution of the PL polarization
\begin{eqnarray}
\nonumber
P(t)&=&\big[I(\sigma_{+})-I(\sigma_{-})\big]/\big[I(\sigma_{+})+I(\sigma_{-})\big]\\
&=&\frac{1}{n_{\rm ex}}\sum_{\bf P}\mbox{Tr}[{\rho({\bf P},t)}I'],
\label{trace}
\end{eqnarray}
with $I(\sigma_{\pm})$ representing the intensity of the $\sigma_{\pm}$ light
and $n_{\rm ex}$=$\sum_{\bf
  P}$Tr[${\rho({\bf P},t)}]$ being the density of the pumped bright excimer;
\begin{equation}
I'=\left(\begin{array}{cccc}
1 & 0 & 0 & 0\\
0 & -1 & 0  & 0 \\
0 & 0  & -1  & 0\\
0 & 0  & 0  & 1\\
\end{array}\right).
\label{hopping2}
\end{equation} 
The initial condition for the density matrix is set to be
\begin{equation}
\rho_{ss}({\bf P},0)=\alpha_{ss}\exp\Big\{-\big[\varepsilon({\bf
  P})-\varepsilon_{\rm pump}\big]^2/(2\Gamma^2)\Big\}
\label{pump}
\end{equation}
and $\rho_{ss'}({\bf P},0)=0$ with $s\ne s'$. Here, $\varepsilon({\bf P})=\hbar^2|{\bf P}|^2/(2m^*)$ is the excimer kinetic energy with
$m^*$ being the excimer effective mass, which is the same as the effective mass of
  the IL and CT excitons; $\varepsilon_{\rm pump}$ is
 the energy of pulse center in reference to the minimum of the excimer energy
 band;
 $\Gamma$ denotes the energy broadening of the pulse;
 \begin{equation}
\alpha_{ss}=\frac{n_{ss}}{\sum_{\bf P}\exp\Big\{-\big[\varepsilon({\bf
    P})-\varepsilon_{\rm pump}\big]^2/(2\Gamma^2)\Big\}},
\label{initial}
\end{equation}
with $n_{\rm ss}$ being the pumped excimer density. In the pump-probe experiment, according to the chiral optical valley
selection rule,\cite{Hamiltonian1} we set $n_{11}=n_{44}=n_{\rm ex}/2$ and $n_{\rm
  22}=n_{\rm 33}=0$ with $P(0)=100\%$. 

\subsubsection{PL depolarization dynamics in the pump-probe setup}
Then we look into the PL depolarization dynamics in the pump-probe setup in the
BL WS$_2$.\cite{WS1} The material parameters in our computation are listed in
Table~\ref{material}.
\begin{table}[htb]
  \caption{Material parameters used in the computation for the KSBEs.}
  \label{material} 
  \begin{tabular}{l l l l}
    \hline
    \hline
    $\kappa_{\parallel}$&\;\;\;\;\;$4.4^a$&$m^*/m_0$&\;\;\;\;\;$0.21^{b}$\\
    $a_{B\parallel}$(nm)&\;\;\;\;\;$0.94$\;\;\;\;&$a_{B\perp}$(nm)&\;\;\;\;\;$0.77$\\
    $n_{\rm ex}$ (cm$^{-2}$)&\;\;\;\;\;$10^{12}$&$\tau_P^*$ (fs)&\;\;\;\;\;$13.0^c$\\
    $\alpha_2$ (${\rm \AA}^2$)&\;\;\;\;\;$4.43$&$\beta(\pm 1)$ (${\rm \AA}^2$)&\;\;\;\;\;$0.14$\\
    $A_{1'}^{\rm IL}$&\;\;\;\;\;$0.79$&$A_{2'}^{\rm CT}$&\;\;\;\;\;$-0.62$\\
    \hline
    \hline
\end{tabular}\\
$^a$ Ref.~\onlinecite{MoS2}. \quad$^b$
 Refs.~\onlinecite{Nano}.
 \quad $^c$ Refs.~\onlinecite{Cui}. 
\end{table}

\begin{figure}[htb]
  {\includegraphics[width=8.2cm]{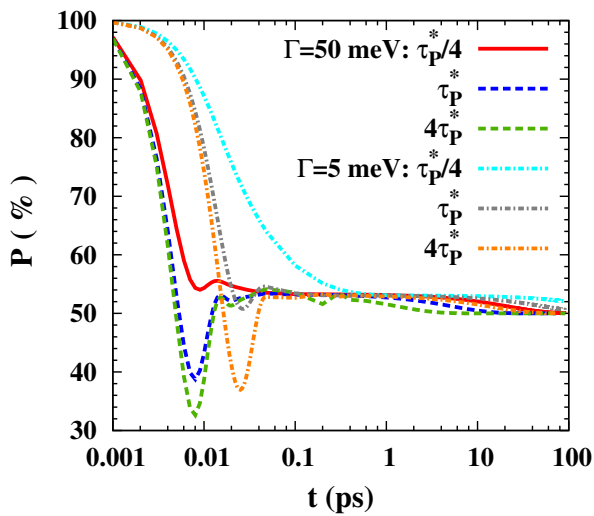}}
  \caption{(Color online) Evolution of the PL polarization when the pumped energy
    centered at the resonance energy for the ${\rm A}'$ bright excimer ($E_1'=1.99$~eV)
 with different energy broadenings ($\Gamma=50$ and 5~meV) 
and momentum relaxation
    times ($\tau^*_P/4$, $\tau^*_P$ and $4\tau^*_P$).}
\label{figyw2}
\end{figure}

 In our computation, as a first step in the investigation, the momentum relaxation time $\tau^*_{P}$
 in Table~\ref{material} is obtained based on the elastic
scattering approximation in the KSBEs.\cite{Sham1}
 Its value is estimated to be 13~fs by considering the measured broadening of the A exciton
 energy $\Gamma\approx 55$~meV at 10~K with $\tau^*_P\approx
 \hbar/\Gamma$.\cite{Cui,WS1} By setting $\varepsilon_{\rm
   pump}=0$~eV in Eq.~(\ref{pump}),\cite{CD,many_body,Korn2}
 with the material parameters in
 Table~\ref{material},
 the evolution of the PL polarization with different energy broadenings and scattering strengths can be
obtained by numerically solving the KSBEs, shown in
Fig.~\ref{figyw2}. 

From Fig.~\ref{figyw2}, several features of the PL depolarization dynamics 
can be obtained. Similar to the experimental results in
 ML TMDs,\cite{CD,many_body} with the absorption of the
$\sigma_{+}$ light, the emergence of the $\sigma_{-}$ light is also 
instantaneous (in the order of 10~fs) when the energy broadening is large ($\Gamma=50$~meV) or/and
the scattering is weak. Moreover, there is also a large residual PL polarization $50\%$, lasting
for extremely long time, which is robust against the initial energy broadening and strength of the
momentum scattering. This seems similar to the situation in ML
TMDs.\cite{CD,many_body,Korn2,Yu3} However, in the BL WS$_2$, the mechanism for the existence of this large residual PL
polarization is different from the ML situation.\cite{CD,many_body,Yu3,Korn2} In ML TMDs,
the residual PL polarization arises from the weak exchange interaction between
the exciton states with $|{\bf P}|\approx 0$, in which the decay of the residual PL polarization
(about $10\%$) is nevertheless
obvious and lasts only for about 10~ps.\cite{CD,many_body,Yu3,Korn2} Moreover,
the residual PL polarization there is sensitive to the experimental conditions and
strength of the scattering. In the following, we show that the anomalous PL
depolarization behavior here, which is very different from the spin relaxation in
semiconductors,\cite{review1,Bloch,Awschalom,Zutic,fabian565,Dyakonov,broadening2,Korn,notebook}
arises from the unique feature of the exchange
interaction [Eq.~(\ref{Coulomb2})] in the BL WS$_2$.

It is interesting to see that when the system evolves into the steady state shown in
  Fig.~\ref{figyw2} with $P=50\%$, the density matrix evolved into has the form
\begin{equation}
\rho_s({\bf P})=a({\bf P})\left(\begin{array}{cccc}
3/8 & 0 & 0 & 1/8\\
0 & 1/8 & -1/8  & 0\\
0& -1/8 & 1/8  & 0\\
1/8 & 0  & 0  & 3/8\\
\end{array}\right).
\label{density}
\end{equation}
Here, $a({\bf P})$ depends on the concrete
initial condition [Eq.~(\ref{pump})] and satisfies the normalized condition $\sum_{\bf P}a({\bf
  P})/n_{ex}=1$. One notes that when the system evolves into this steady
  state [Eq.~(\ref{density})] with the initial condition $n_{11}=n_{44}=n_{\rm ex}/2$ and
    $n_{22}=n_{33}=0$, $(n_{11}+n_{44})/(n_{22}+n_{33})=3:1$ is satisfied and $P=\frac{1}{n_{ex}}\sum_{\bf P}\mbox{Tr}[\rho_s({\bf
   P})I']$ is calculated to be $50\%$ exactly. Furthermore, when the system is polarized by the
    $\sigma_{-}$ light with the initial condition being $n_{11}=n_{44}=0$ and
    $n_{22}=n_{33}=n_{\rm ex}/2$, we
    find $(n_{11}+n_{44})/(n_{22}+n_{33})=1:3$ and
    $P=-50\%$ in the steady state.
 Moreover, it is easy to verify
 that this density
 matrix commutes with the exchange interaction Hamiltonian
 [Eq.~(\ref{Coulomb2})]
\begin{equation}
\big[H_{\rm ex}({\bf P}), \rho_s({\bf P})\big]=0,
\end{equation}
with
 $\rho_s({\bf P})H_{\rm ex}({\bf P})=H_{\rm ex}({\bf
    P})\rho_{s}({\bf P})=a({\bf P}) H_{\rm ex}({\bf P})/4$. 
Hence, from the KSBEs
  [Eq.~{\ref{ksbe}}], this guarantees the residual PL polarization is in the
  steady state. 

In Appendix~\ref{BB}, we extend our formula to the situation
  with the system pumped by the elliptically polarized
light analytically. With the polarization of the elliptically polarized light being
$x=\big[I(\sigma_{+})-I(\sigma_{-})\big]/\big[I(\sigma_{+})+I(\sigma_{-})\big]$, 
we show that the residual PL polarization is always $x/2$,
 which is half of the initial polarization of the elliptically polarized
light. Furthermore, the steady-state density matrix $\rho_{s}({\bf P})$ is
proved to be
\begin{equation}
\rho_{s}({\bf P})=\frac{a({\bf P})}{4}\left(\begin{array}{cccc}
1+x/2 & 0 & 0 & x/2\\
0 & 1-x/2 & -x/2  & 0 \\
0 & -x/2  & 1-x/2  & 0\\
x/2 & 0  & 0  & 1+x/2\\
\end{array}\right).
\label{initial4}
\end{equation} 
Therefore, the steady-state density matrix Eq.~(\ref{density}) with the system
pumped by the $\sigma_{+}$ light ($x=100\%$) is only a special situation in Eq.~(\ref{initial4}).

Finally, we address the recent steady-state measurement of the PL polarization
by Zhu {\em et al.} for the BL WS$_2$, in which the anomalous PL
polarization as large as $P\approx 100\%$ is observed.\cite{Cui} The puzzle of
the experiment is that under the same experimental
condition, the measured PL polarization in the BL WS$_2$ is anomalously larger than
the ML situation, in which $P$ is less than
$40\%$,\cite{Cui} and hence this cannot be understood by the
spin-layer locking picture by Jones {\em et al.}.\cite{Hamiltonian1,Hamiltonian2}
 However, this experiment can be well understood according to our calculation in the
 BL WS$_2$ based on the exchange interaction. In the BL WS$_2$, the PL relaxation
time $\tau_s$ should be extremely long when the system is at the steady state.
 Moreover, according to the rate equation,\cite{absorption_Mak,
absorption_Kioseoglou,valley_wang18,absorption_Marie} the steady-state PL
polarization is
derived to be  
 \begin{equation}
P=P_0/(1+2\tau_r/\tau_s),
\label{reasonable}
\end{equation}    
with $P_0\approx 100\%$ being the inital PL polarization. Hence when $\tau_s\gg
\tau_r$, we obtain $P\approx 100\%$. Whereas
in the ML WS$_2$, it has been well understood that the e-h exchange interaction can cause PL
depolarization efficiently.\cite{Yu3,Glazov}

\section{Conclusion and discussion}
\label{summary}
In conclusion, we have investigated the excimer 
  excitation spectra and the PL 
depolarization dynamics in BL WS$_2$. We first present a possible understanding for the optical
  excitation spectra for the recent PL experiments by Zhu {\em
    et al.} in the BL WS$_2$,\cite{Cui,WS1} in which four resonance excitations (1.68, 1.93, 1.99 and
2.37~eV) are observed and speculated to be the indirect exciton for the $\Gamma$ valley,
trion, A exciton
and B exciton excitations, respectively.\cite{Cui,WS1} Furthermore,
 in the experiment,\cite{Cui,WS1} the redshift for the A exciton energy about 30 $\sim$ 50~meV is observed when the sample is synthesized from ML to BL.
In our study, we find that in the BL WS$_2$,
 due to the efficient inter-layer hopping of the hole, the excimer states can
be formed from the superposition of the IL and CT
 excitons.\cite{dielectric_B,directgap_wang23,dielectric,dielectric_c}  According
 to our study, the energy levels of the four experimentally observed optical excitations
in the BL WS$_2$ are calculated to be 1.69, 1.99, 2.10 and 2.41~eV,
  corresponding to A CT exciton, ${\rm A}'$ excimer, ${\rm B}'$ excimer and B IL
exciton,
respectively. Here, the ${\rm A}'$ (${\rm B}'$) excimer state is composed
of the A IL
and B CT exciton states. These calculations show good agreement with
the recent experiments by Zhu {\em
    et al.},\cite{Cui,WS1} but with different understanding for the 
  first three
  elementary excitations.  Furthermore, the binding energy for the ${\rm A}'$ excimer state is calculated
to be 40 meV, in consistent with 30 $\sim$ 50~meV 
observed in the experiment.\cite{Cui,WS1} Based on the excimer
state, we further derive the e-h exchange interaction including
 all the dominant processes. With the transition channel between dark excitons
 forbidden,\cite{Yu3,Glazov} we find both the intra-
  and inter-valley exchange interactions can cause the bright excimer transition
  due to the
MSS mechanism.\cite{Sham1,Sham2}

We then study the PL depolarization dynamics due to the e-h
exchange interaction in the pump-probe setup based on the KSBEs. We find that with the absorption of the
$\sigma_{+}$ light, the emergence of the $\sigma_{-}$ light can be instantaneous,
which is similar to the ML situation.\cite{CD,many_body,Yu3,Korn2} Moreover, we further find that 
there is always a residual PL polarization as large as $50\%$, lasting
for extremely long time, which is robust against the initial energy broadening and strength of the
momentum scattering. This large steady-state PL polarization indicates that the
PL relaxation time is extremely long in the BL WS$_2$ in the steady state and can be the cause of the
anomalously large PL polarization nearly $100\%$ observed in the
experiment by Zhu {\em
    et al.} in the BL WS$_2$.\cite{Cui} 
This steady state is shown to come from the unique form of the 
exchange interaction
  Hamiltonian [Eq.~(\ref{Coulomb2})], under which the density matrix evolves
  into the state $\rho_s({\bf P})$ [Eq.~(\ref{density})] which 
communicates with the exchange interaction Hamiltonian 
$\big[H_{\rm ex}({\bf P}), \rho_s({\bf P})\big]=0$, with
 $\rho_s({\bf P})H_{\rm ex}({\bf P})=H_{\rm ex}({\bf
    P})\rho_{s}({\bf P})\propto H_{\rm ex}({\bf P})$. Specifically, from
    the density matrix $\rho_s({\bf P})$ [Eq.~(\ref{density})], one further
    observes that when the system is polarized by the
    $\sigma_{+}$ light, in the steady state, the density ratio of the bright
    excimers associated with the $\sigma_{+}$ and $\sigma_{-}$ light is $3:1$; 
whereas when
   the system is polarized by the $\sigma_{-}$ light, this ratio is $1:3$.
 Furthermore, in general, if the system is pumped by the elliptically polarized
light, we have demonstrated that the residual
  PL polarization is always half of the initial polarization of the elliptically polarized
light.
Moreover, it is 
 noted that although this specific exchange interaction
 Hamiltonian [Eq.~(\ref{Coulomb2})] is
derived based on the excimer states, its contribution is mainly from the
 exchange interaction between the two IL excitons
 [Eq.~(\ref{Coulomb})]. 
  
It should be noted that rather than our approach by dealing first with the
  strong Coulomb interaction and then the influence of the inter-layer hopping of the hole for
  the excimer state,\cite{exchange,Sham1,exchange,dot,Yu3} Jones {\em et al.} presented other treatment for the
exciton states in BL TMDs by considering the influence of the inter-layer hopping of the
hole.\cite{Hamiltonian2} In their treatment, Jones {\em et al.} first diagonalize the ${{\bf k}\cdot{\bf p}}$
Hamiltonian [Eq.~(\ref{Hamiltonian})] and then construct the exciton states with
the eigenstates of the Hamiltonian.\cite{Hamiltonian1,Hamiltonian2} However, due
to the fact that 
the inter-layer hopping energy for the hole is smaller than the valence bands
energy 
splitting, the mixture of the wavefunction of the holes in different layers is
negligible and hence the BL TMDs
can be treated as two separated ML TMDs, which is referred to as the spin-layer locking
effect in their study.\cite{Hamiltonian1,Hamiltonian2} However, this treatment is correct only when the
strength of the 
Coulomb interaction is weak, and hence the exciton binding energy is much smaller than the inter-layer
hopping energy for the hole. Whereas in BL TMDs, the experimentally measured exciton
binding energy is much larger than the inter-layer hopping energy for the
hole.\cite{MoS2,WS1,WS2,WS3,MoSe2,WSe2,Hamiltonian1,Hamiltonian2} Therefore, one should first deal with the
Coulomb interaction and then the effect of the inter-layer hopping of the hole to get the
correct picture for the excitation in BL TMDs, as we do in this study.

 Furthermore, there also exists other speculation for the lowest excitation
   with excitation energy $E\approx 1.68$ eV observed in the BL WS$_2$ in the
   literature.\cite{Cui,indirect} Zhao {\em et al.}\cite{indirect} and Zhu {\em
     et al.}\cite{Cui} claimed that this excitation comes from the indirect
   excitation for the $\Gamma$ valley.
 This claim is still controversial. On one hand, this is in contrast to the
 understanding in the BL TMD
heterostructures, where the lowest excitation is considered to be the CT
exciton.\cite{long_live,inter-layer} On the other hand, the indirect excitation
needs to involve 
a high-order phonon absorption/emission process, whose efficiency can be very
low in the optical process. According to our calculation, this excitation
  comes from the CT exciton. More investigations are
needed to further clarify this problem. 
   
 Finally, we summarize the several approximations in our study. First, the excimer
  excitation spectra are calculated from the material parameters constructed from
  the ML TMDs including
  experimental measurements\cite{MoS2,WS1,WS2,WS3,MoSe2,WSe2} and theoretical
 calculations.\cite{Hamiltonian1,Hamiltonian2,dielectric}
 However, when the sample is synthesized from ML to BL, both the energy bands and
 the dielectric environment can be influenced.\cite{indirect,dielectric,WS1} This may also cause the
 energy shift for the optical excitation in BL TMDs compared to the ML
 situation. Nevertheless,
 this cannot modify the physical picture for the four elementary excitations we
 reveal here. Second, in our calculation, we only include the bright exciton for
 the excimer excitation energy. This is because although the dark exciton can also
   contribute to the formation of the excimer state, on one hand, it has negligible influence
   on the excimer energy level; on the other hand, it cannot be excited in the optical
   process. Hence, in the optical
   process, only the bright exciton is 
   considered.

\begin{acknowledgments}
This work was supported
 by the National Natural Science Foundation of China under Grant
No. 11334014, the National Basic Research Program of China under Grant No.
2012CB922002 and the Strategic Priority Research Program 
of the Chinese Academy of Sciences under Grant
No. XDB01000000.  
\end{acknowledgments}

 \begin{appendix}
 \section{Excimer Hamiltonian}
 \label{AA}
 In this appendix, based on the ${\bf
   k}\cdot {\bf p}$ Hamiltonian, we give the explicit
 form of the 
 excimer Hamiltonian $H^{\rm eh}_{m'n'\atop mn}{{\bf r}_1'~{\bf r}_2'\choose {\bf r}_1~{\bf
    r}_2}$ for the exciton envelop function $\tilde{F}_{mn}({\bf r}_1,{\bf
    r}_2)$ in the coordinate space, where $m (m')$ and $n (n')$ denote the 
  indices including the layer, valley and spin degrees of freedom
 for the electron and hole.\cite{exchange,Sham1,dot}

The ${\bf
   k}\cdot {\bf p}$ Hamiltonian with the basis $|d^u_{z^2}\rangle$, $|d^l_{z^2}\rangle$,
$\frac{1}{\sqrt{2}}(|d^u_{x^2-y^2}\rangle-i\tau_z|d^u_{xy}\rangle)$ and 
 $\frac{1}{\sqrt{2}}(|d^l_{x^2-y^2}\rangle+i\tau_z|d^l_{xy}\rangle)$ 
reads\cite{Hamiltonian1,Hamiltonian2}
\begin{widetext}
\begin{equation}
\hat{H}=\left(\begin{array}{cccc}
\Delta-\tau_zs_z\lambda_c+Ed/2 & 0 & at(\tau_zk_x+ik_y) & 0\\
0 & \Delta+\tau_zs_z\lambda_c-Ed/2 & 0 & at(\tau_zk_x-ik_y)\\
at(\tau_zk_x-ik_y) & 0 & -\tau_zs_z\lambda_v+Ed/2 & \tau_{\perp}\\
0 & at(\tau_zk_x+ik_y) & \tau_{\perp} & \tau_zs_z\lambda_v-Ed/2
\end{array}\right).
\label{Hamiltonian}
\end{equation}
\end{widetext}
Here, $a$ is the lattice constant and $t$ represents the effective
hopping integral; $\Delta$ is the band gap; $2\lambda_c$ ($2\lambda_v$) represents the energy
  splitting for the conduction (valence) bands; $\tau_{\perp}$ denotes the
  inter-layer hopping
 for the hole (it vanishes for the 
  electron); $\tau_z=\pm 1$ stands for the valley index with $\tau_z=1$ ($-1$)
  for the K (K$'$) valley; 
$s_z$ denotes the Pauli spin matrix;  $E$ and
$d$
 are the magnitude of the electric field
and the interlayer distance, respectively.

The eigenequation
  expressed by the excimer Hamiltonian for the exciton envelop function
  satisfies
\begin{equation}
\sum_{mn}\int d{\bf r}_1 d{\bf r}_2 H^{\rm eh}_{m'n'\atop mn}{{\bf r}_1'~{\bf r}_2'\choose {\bf r}_1~{\bf
    r}_2}\tilde{F}_{mn}({\bf r}_1,{\bf r}_2)=E\tilde{F}_{m'n'}({\bf r}'_1,{\bf r}'_2),
\end{equation}
where
\begin{eqnarray}
\nonumber
&&H^{\rm eh}_{m'n'\atop mn}{{\bf r}_1'~{\bf r}_2'\choose {\bf r}_1~{\bf
    r}_2}=\big[H^e_{m'm}({\bf k}_1)\delta_{n'n}+H^h_{n'n}({\bf
  k}_2)\delta_{m'm}\\
\nonumber
&&\mbox{}+U^{\rm eh}({\bf r}_1-{\bf
  r}_2)\delta_{m'm}\delta_{n'n}+T_{m'n'\atop mn}\big]\delta({\bf r}_1-{\bf
  r}'_1)\delta({\bf r}_2-{\bf r}'_2)\\
&&\mbox{}+U^{\rm ex(1)}_{m'n'\atop mn}{{\bf r}_1'~{\bf r}_2'\choose {\bf r}_1~{\bf
    r}_2}+U^{\rm ex(2)}_{m'n'\atop mn}{{\bf r}_1'~{\bf r}_2'\choose {\bf r}_1~{\bf
    r}_2}.
\label{effective_H}
\end{eqnarray}
Here, ${\bf k}=-i\nabla$, 
\begin{equation}
U^{\rm eh}({\bf r}_1-{\bf r}_2)=-\frac{e^2}{4\pi \varepsilon_0\kappa_l|{\bf
    r}_1-{\bf r}_2|},
\end{equation}
with $\kappa_l$ being $\kappa_{\parallel}$
($\kappa\equiv\sqrt{\kappa_{\parallel}\kappa_{\perp}}$) if the electron in
the $m$-band and hole in the $n$-band
are in the same (different) layer;
\begin{eqnarray}
\nonumber
&&H_{m'm}^e({\bf k}_1)=E_m({\bf
  k}_0)\delta_{m'm}~~~~~~\\
\nonumber
&&\mbox{}+\frac{\hbar^2}{2m_0^2}\sum_{m''}\big[{\bf k}_1\cdot {\bgreek
  \pi}_{m'm''}({\bf k}_0)\big]\big[{\bf k}_1\cdot {\bgreek \pi}_{m''m}({\bf
  k}_0)\big]~~~~~~\\
&&\times\Big[\frac{1}{E_m({\bf k}_0)-E_{m''}({\bf
    k}_0)}+\frac{1}{E_{m'}({\bf k}_0)-E_{m''}({\bf k}_0)}\Big];~~~~~~
\label{ee}
\end{eqnarray}
\begin{equation}
H^h_{n'n}({\bf k}_2)=-H^e_{\Theta n \Theta n'}(-{\bf k}_2);
\end{equation}
and
\begin{equation}
T_{m'n'\atop mn}=T_{n'n}\delta_{m'm}
\end{equation}
with $T_{n'n}$ being nonzero $\tau_{\perp}$) only when the holes in the $n$-
  and $n'$-bands are located in the different layers
  with the same valley and spin degrees of freedom.
In Eq.~(\ref{ee}), ${\bgreek \pi}={\bf p}+\frac{\hbar}{4m_0^2c^2}[{\bgreek
  \sigma}\times (\nabla V_0)]$ with $V_0$ denoting the lattice
potential. ${\bgreek \pi}_{{\eta}{\eta}'}({\bf k}_0)$ stands for the matrix elements of $\bgreek \pi$
between two Bloch wavefunctions with indices $\eta$ and ${\eta}'$.
 The nonzero expressions of ${\bgreek \pi}_{{\eta}{\eta}'}({\bf k}_0)$
can be obtained from the Hamiltonian Eq.~({\ref{Hamiltonian}}).
For the K ($\tau_z=1$) and K$'$ ($\tau_z=-1$) valleys,
\begin{eqnarray}
\nonumber
\langle \uparrow_c^u|\pi_x|\uparrow_v^u\rangle&=&\langle
\downarrow_c^u|\pi_x|\downarrow_v^u\rangle=\tau_z m_0at/\hbar,\\
\nonumber
\langle \uparrow_c^u|\pi_y|\uparrow_v^u\rangle&=&\langle
\downarrow_c^u|\pi_y|\downarrow_v^u\rangle=i m_0at/\hbar,\\
\nonumber
\langle \uparrow_c^l|\pi_x|\uparrow_v^l\rangle&=&\langle
\downarrow_c^l|\pi_x|\downarrow_v^l\rangle=\tau_z m_0at/\hbar,\\
\langle \uparrow_c^l|\pi_y|\uparrow_v^l\rangle&=&\langle
\downarrow_c^l|\pi_y|\downarrow_v^l\rangle=-i m_0at/\hbar.
\end{eqnarray}

We then express the e-h exchange interaction. For $U^{\rm ex(1)}_{m'n'\atop mn}{{\bf r}_1'~{\bf r}_2'\choose {\bf r}_1~{\bf
    r}_2}$, it describes that the e-h pair in one IL exciton can virtually recombine
and then generate another IL exciton due to the Coulomb interaction
directly. We express the e-h exchange interaction Hamiltonian for both the L-R and S-R
parts:
\begin{equation}
U^{\rm ex(1)}_{m'n'\atop mn}{{\bf r}_1'~{\bf r}_2'\choose {\bf r}_1~{\bf
    r}_2}=H^{\rm LR}_{m'n'\atop mn}{{\bf r}_1'~{\bf r}_2'\choose {\bf r}_1~{\bf
    r}_2}+H^{\rm SR}_{m'n'\atop mn}{{\bf r}_1'~{\bf r}_2'\choose {\bf r}_1~{\bf
    r}_2}.
\end{equation} 
For the L-R part,
\begin{eqnarray}
\nonumber
&&H^{\rm LR}_{m'n'\atop mn}{{\bf r}_1'~{\bf r}_2'\choose {\bf r}_1~{\bf
    r}_2}=-\sum_{\alpha\beta}\frac{\hbar^2}{2m_0^2}\pi_{\Theta
n m}^{\alpha}({\bf k}_0)\pi_{m'\Theta n'}^{\beta}({\bf k}_0')\\
\nonumber
&&\mbox{}\times\Big\{\frac{1}{\big[E_m({\bf
    k}_0)-E_n({\bf k}_0)\big]^2}+\frac{1}{\big[E_{m'}({\bf k}_0')-E_{n'}({\bf k}_0')\big]^2}\Big\}\\
&&\mbox{}\times\frac{\partial^2}{\partial {\bf
  r}_1^{\alpha}\partial {\bf r}_1^{\beta}}U({\bf r}_1-{\bf r}'_2)\delta({\bf r}_1-{\bf r}_2)\delta({\bf r}'_1-{\bf r}'_2),
\end{eqnarray}
with $\alpha$ ($\beta$) denoting $x$ or $y$.
For the S-R part, 
\begin{equation}
H^{\rm SR}_{m'n'\atop mn}{{\bf r}_1'~{\bf r}_2'\choose {\bf r}_1~{\bf
    r}_2}=SU_{m' \Theta n \atop \Theta n'
  m}\delta({\bf r}_1-{\bf r}_2)\delta({\bf r}_1-{\bf r}_1')\delta({\bf r}_2-{\bf
r}_2'),
\label{SR}
\end{equation}
with 
\begin{eqnarray}
\nonumber
&&U_{m' \Theta n \atop \Theta n'm}=\frac{1}{S^2}\int d{\bf r}_1d{\bf r}_2
\big[\Psi^{m'}_{{\bf k}_0'}({\bf r}_1)\big]^*\big[\Theta\tilde{\Psi}^n_{{\bf k}_0}({\bf
  r}_2)\big]^*\\
&&\mbox{}\times U({\bf r}_1-{\bf r}_2)\big[\Theta \tilde{\Psi}^{n'}_{{\bf k}_0'}({\bf
  r}_1)\big]\Psi^m_{{\bf k}_0}({\bf r}_2).
\end{eqnarray}
Here, $S$ is the area of the 2D plane of the BL WS$_2$. 

For $U^{\rm ex(2)}_{m'n'\atop mn}{{\bf r}_1'~{\bf r}_2'\choose {\bf r}_1~{\bf
    r}_2}$, it describes that the hole in the CT exciton first hops from one layer to another
and then recombines virtually with the electron part to generate the IL
exciton
 due to the Coulomb interaction.
 The dominant process of this exchange interaction is the L-R part, which is written as 
\begin{eqnarray}
\nonumber
&&U^{\rm ex(2)}_{m'n'\atop mn}{{\bf r}_1'~{\bf r}_2'\choose {\bf r}_1~{\bf
    r}_2}\\
\nonumber
&&=-\frac{\hbar^2}{2m_0^2}\sum_{\alpha\beta}\sum_{l''}\frac{\pi_{m'\Theta
    n'}^{\alpha}({\bf k}_0)\pi_{\Theta l''
    m}^{\beta}({\bf k}_0')T_{\Theta n \Theta
    l''}}{[E_m({\bf k}_0)-E_{l''}({\bf k}_0)][E_m({\bf k}_0)-E_n({\bf k}_0)]}\\
\nonumber
&&\mbox{}\times\Big[\frac{1}{E_m({\bf k}_0)-E_{l''}({\bf k}_0)}+\frac{1}{E_m({\bf k}_0)-E_n({\bf k}_0)}\Big]\\
&&\mbox{}\times\frac{\partial^2}{\partial {\bf
  r}_1^{\alpha}\partial {\bf r}_1^{\beta}}U({\bf r}_1-{\bf r}'_2)\delta({\bf r}_1-{\bf r}_2)\delta({\bf r}'_1-{\bf r}'_2).
\end{eqnarray}

 \section{Derivation of the steady-state density matrix $\rho_s({\bf P})$}
\label{BB}
In this appendix, we derive the steady-state density matrix $\rho_s({\bf P})$ 
based on the KSBEs [Eq.~(\ref{ksbe})]. Generally, the
system can be initialized by the elliptically polarized light with the
polarization being
$x=\big[I(\sigma_{+})-I(\sigma_{-})\big]/\big[I(\sigma_{+})+I(\sigma_{-})\big]$
which varies from $-100\%$ to $100\%$. Accordingly, the pumped electron
density associated with the $\sigma_{+}$ ($\sigma_{-}$) light is $n_{\rm ex}(1+x)/2$
[$n_{\rm ex}(1-x)/2$] and hence from Eqs.~(\ref{pump}) and (\ref{initial}), the
initial density matrix for the system can be written as
\begin{equation}
\rho_{i}({\bf P})=\frac{a({\bf P})}{4}\left(\begin{array}{cccc}
1+x & 0 & 0 & 0\\
0 & 1-x & 0  & 0 \\
0 & 0  & 1-x  & 0\\
0 & 0  & 0  & 1+x\\
\end{array}\right),
\label{initial2}
\end{equation}
with 
\begin{equation}
a({\bf P})=\frac{\exp\Big\{-\big[\varepsilon({\bf
  P})-\varepsilon_{\rm pump}\big]^2/(2\Gamma^2)\Big\}}{\sum_{\bf P}\exp\Big\{-\big[\varepsilon({\bf
    P})-\varepsilon_{\rm pump}\big]^2/(2\Gamma^2)\Big\}}.
\end{equation}
In the following, we demonstrate that with this initial condition [Eq.~(\ref{initial2})], the
system evolves into the steady state with residual polarization $P(t)=x/2$,
 and the corresponding steady-state density matrix $\rho_s({\bf P})$ is presented. 

The exchange interaction Hamiltonian can be splitted
into the off-block-diagonal [$H_{\rm ex}^{(1)}({\bf P})$] and block-diagonal [$H_{\rm ex}^{(2)}({\bf P})$]\cite{block} parts
\begin{equation}
H_{\rm ex}({\bf P})=H_{\rm ex}^{(1)}({\bf P})+H_{\rm ex}^{(2)}({\bf P}),
\end{equation}
with 
\begin{equation}
H_{\rm ex}^{(1)}({\bf P})=\left(\begin{array}{cccc}
0 & P_{+}^2 & -P_{+}^2 & 0\\
P_{-}^2 & 0 & 0  & -P_{-}^2 \\
-P_{-}^2 & 0  & 0  & P_{-}^2\\
0 & -P_{+}^2  & P_{+}^2  & 0\\
\end{array}\right)
\end{equation}
 and 
\begin{equation}
H_{\rm ex}^{(2)}({\bf P})=\left(\begin{array}{cccc}
|{\bf P}|^2 & 0 & 0 & -|{\bf P}|^2\\
0 & |{\bf P}|^2 & -|{\bf P}|^2  & 0 \\
0 & -|{\bf P}|^2  & |{\bf P}|^2  & 0\\
-|{\bf P}|^2 & 0  & 0  & |{\bf P}|^2\\
\end{array}\right).
\end{equation}
It can be shown that the block-diagonal part [$H_{\rm ex}^{(2)}({\bf P})$] in the exchange
interaction Hamiltonian [Eq.~(\ref{Coulomb2})] has no effect on the PL
depolarization dynamics. 
From the KSBEs [Eq.~(\ref{ksbe})], the depolarization dynamics for $P_0({\bf
  P},t)=\mbox{Tr}[{\rho({\bf P},t)}I']$ can be written as
\begin{eqnarray}
\nonumber
&&\partial_t{\mbox{Tr}[\rho({\bf P},t) I']}+i\mbox{Tr}[H_{\rm
ex}({\bf P})\rho({\bf P},t) I']/\hbar\\
&&\mbox{}-i\mbox{Tr}[\rho({\bf P},t) H_{\rm ex}({\bf P}) I']/\hbar=0.
\label{ksbe2}
\end{eqnarray}
Therefore, from Eq.~(\ref{ksbe2}), with $H_{\rm ex}^{(2)}({\bf P})I'=I'H_{\rm ex}^{(2)}({\bf P})$, $\mbox{Tr}[H^{(2)}_{\rm
ex}({\bf P})\rho({\bf P},t) I']=\mbox{Tr}[\rho({\bf P},t) H^{(2)}_{\rm ex}({\bf P})
I']$ and hence $H^{(2)}_{\rm ex}({\bf P})$ has no
effect on the PL depolarization dynamics. 

Accordingly, with the off-block-diagonal part [$H^{(1)}_{\rm ex}({\bf
  P})$] of the exchange interaction
Hamiltonian,
 for the system in the steady state, the condition $[H^{(1)}_{\rm ex}({\bf P}),\rho({\bf
  P},t)]=0$ for any ${\bf P}$ is satisfied. Hence, for the density matrix 
\begin{equation}
\rho({\bf P},t)=\left(\begin{array}{cccc}
\rho_{11}({\bf P},t) & \rho_{12}({\bf P},t) & \rho_{13}({\bf P},t) &\rho_{14}({\bf P},t)\\
 & \rho_{22}({\bf P},t) & \rho_{23}({\bf P},t)  & \rho_{24}({\bf P},t) \\
 &   & \rho_{33}({\bf P},t)  & \rho_{34}({\bf P},t)\\
 &   &   & \rho_{44}({\bf P},t)\\
\end{array}\right),
\label{density_matrix}
\end{equation} 
it can be proved that
\begin{equation}
\left\{\begin{array}{cc}
\rho_{11}({\bf P},t)=\rho_{44}({\bf P},t)~~~~~~~~~~~~~~~~~~~~~~~~~~~~~~\\
\rho_{22}({\bf P},t)=\rho_{33}({\bf P},t)~~~~~~~~~~~~~~~~~~~~~~~~~~~~~~\\
\rho_{11}({\bf P},t)-\rho_{22}({\bf P},t)=\rho_{14}({\bf
  P},t)-\rho_{23}({\bf P},t)\\
\rho_{14}^*({\bf P},t)=\rho_{14}({\bf P},t)~~~~~~~~~~~~~~~~~~~~~~~~~~~~~~\\
\rho_{23}^*({\bf P},t)=\rho_{23}({\bf P},t)~~~~~~~~~~~~~~~~~~~~~~~~~~~~~~
\end{array}\right.,
\label{condition1}
\end{equation} 
and the other matrix elements are zero. 

Then with the conditions [Eq.~(\ref{condition1})], in order to obtain the exact values
of the nonzero terms in the density matrix Eq.~(\ref{density_matrix}), we can
derive the relations between these terms based on the dynamical evolution of
 the density matrix with the initial condition Eq.~(\ref{initial2}). The
 dynamical evolution of the density matrix without the scattering can be obtained with
 the Baker-Hausdorff formula, which reads
\begin{eqnarray}
\nonumber
\rho({\bf P},t)&=&\exp[i H_{\rm ex}^{(1)}({\bf P})t]\rho_i({\bf P})\exp[-i
H_{\rm ex}^{(1)}({\bf P})t]\\
\nonumber
\mbox{}&=&\rho_i({\bf P})+it[H_{\rm ex}^{(1)}({\bf P}),\rho_i({\bf
  P})]\\
\nonumber
\mbox{}&+&\frac{i^2t^2}{2!}\big[H_{\rm ex}^{(1)}({\bf P}),[H_{\rm ex}^{(1)}({\bf
  P}),\rho_i({\bf P})]\big]+\cdots\\
\mbox{}&+&\frac{i^nt^n}{n!}H^{(n)}_{\rm com}({\bf P})+\cdots,
\end{eqnarray} 
with 
\begin{equation}
H^{(n)}_{\rm com}({\bf P})=\Big[\underbrace{H_{\rm ex}^{(1)}({\bf P}),\big[H_{\rm ex}^{(1)}({\bf P}),\cdots[H_{\rm ex}^{(1)}({\bf
  P})}_{n},\rho_i({\bf P})]\big]\Big]
\end{equation} 
for $n=1,2,3\cdots$.

It can be calculated that
\begin{eqnarray}
\nonumber
&&H^{(\rm 2n-1)}_{\rm com}({\bf P})=\frac{1}{2}\big(16|{\bf P}|^4\big)^{\rm n-1}xa({\bf P})\\
&&\times\left(\begin{array}{cccc}
0 & -P_{+}^2 & P_{+}^2 & 0\\
P_{-}^2 & 0 & 0  & -P_{-}^2 \\
-P_{-}^2 & 0  & 0  & P_{-}^2\\
0 & P_{+}^2  & -P_{+}^2  & 0\\
\end{array}\right),
\end{eqnarray}
and
\begin{eqnarray}
\nonumber
&&H^{(\rm 2n)}_{\rm com}({\bf P})=2\big(16|{\bf P}|^4\big)^{\rm n-1}xa({\bf P})\\
&&\times\left(\begin{array}{cccc}
|{\bf P}|^4 & 0 & 0 & -|{\bf P}|^4\\
0 & -|{\bf P}|^4 & |{\bf P}|^4  & 0 \\
0 & |{\bf P}|^4  & -|{\bf P}|^4  & 0\\
-|{\bf P}|^4 & 0  & 0  & |{\bf P}|^4\\
\end{array}\right).
\label{principle}
\end{eqnarray}
From Eq.~(\ref{principle}), one concludes that 
\begin{equation}
\left\{\begin{array}{cc}
\rho_{11}({\bf P},t)+\rho_{14}({\bf P},t)=a({\bf P})(1+x)/4\\
\rho_{22}({\bf P},t)-\rho_{14}({\bf P},t)=a({\bf P})(1-x)/4\\
\rho_{14}({\bf P},t)=-\rho_{23}({\bf P},t)~~~~~~~~~~~~~~~~~~~~
\end{array}\right..
\label{condition2}
\end{equation} 

From Eqs.~(\ref{condition1}) and (\ref{condition2}), when the system is in the
steady state, we obtain 
\begin{equation}
\rho_{s}({\bf P})=\frac{a({\bf P})}{4}\left(\begin{array}{cccc}
1+x/2 & 0 & 0 & x/2\\
0 & 1-x/2 & -x/2  & 0 \\
0 & -x/2  & 1-x/2  & 0\\
x/2 & 0  & 0  & 1+x/2\\
\end{array}\right).
\label{initial3}
\end{equation} 
Therefore, from
Eq.~(\ref{trace}), the steady-state PL polarization is calculated to be
$P(t)=x/2$, which is half of the polarization of the elliptically polarized
light. Specifically, with the system pumped by the $\sigma_{+}$ ($\sigma_{-}$)
light, $x=100\%$ ($x=-100\%$) and hence the
steady-state PL polarization is $P(t)=50\%$ [$P(t)=-50\%$] exactly.
\end{appendix}

\end{document}